\documentclass{article}
\usepackage{hiph-preprint}
\volnumber{22} \issuenumber{1} \edyear{2005}                             
\frompage{000} \topage{000}                                              
\recrevdate{1 January 2005}                                              
\def\rightmark{Freeze out in narrow and wide layers}  
\def\leftmark{V.K. Magas et al.}

\def\be{\begin{equation}}
\def\ee{\end{equation}}
\def\bea{\begin{eqnarray}}
\def\eea{\end{eqnarray}}

\title{Freeze out in narrow and wide layers}
\authors{
{V.K. Magas$^{1}$, L.P. Csernai$^{2}$ and E. Moln\'ar$^{2}$ %
\index{V.K. Magas et al.} 
}\\[2.812mm]
{\normalsize
\hspace*{-8pt}$^1$~Departament d'Estructura i Constituents de la Mat\'eria, \\
Universitat de Barcelona, Diagonal 647, 08028 Barcelona, Spain\\[0.2ex]
\hspace*{-8pt}$^2$ Section for Theoretical and Computational Physics,\\
     University of Bergen, Allegaten 55, 5007 Bergen, Norway
}}

\abstract{The freeze out of particles from a layer of finite 
thickness is discussed in a phenomenological kinetic model. The proposed model,
based on the Modified Boltzman Transport Equation, is Lorentz invariant and 
can be applied equally well for the freeze out layers with space-like and 
time-like normal vectors. It leads to non-equilibrated post freeze out distributions. 
The dependence of the resulting distribution on the thickness of the layer is presented
and discussed for a space-like freeze out scenario.}
\keyword{freeze out, kinetic model}

\PACS{51.10.+y}

\makeindex
\begin{document}

\maketitle

The hydrodynamical modeling of relativistic heavy ion collisions includes three main stages:
the initial stage, the fluid-dynamical stage and the so-called Freeze Out (FO) stage,
when the hydrodynamical description breaks down.
The FO stage is essentially the last part of a collision process and the main source for observables.

Nowadays, FO is usually simulated in two extreme ways: FO happens on the hypersurface with zero thickness, or FO simulated by volume emission model, which in principle requires an infinite time and space for a complete FO. For practical purposes one has to choose between these two, in some sense opposite, scenarios, realized in many different specific models, because it is really very problematic to model FO from the first principles. As it was shown recently in Refs. \cite{ModifiedBTE} during FO system has to be described by the Modified Boltzman Transport Equation (Modified BTE), which is even more complicated than ordinary BTE. 
When the characteristic length scale, describing the change of the distribution function, becomes smaller than mean free path (this always happens at late stages of the FO), then the basic assumptions of the BTE get violated, and the expression for the collision integral has to be modified to follow the trajectories of colliding particles. One can avoid troubles with FO modeling using hydro+cascade two module model \cite{hydro_cascade}, since in hadron cascade models gradual FO is realized automatically. In fact, in cascade model one follows the trajectories of the colliding particles, therefore what is effectively solved is not BTE, but Modified BTE!

Once the necessity of the Modified BTE and at the same time the difficulty of its direct solution  were
realized, the simplified  kinetic FO models become even more important for the understanding of
the principal features of these phenomenon.
In the recent papers \cite{Mo05a,Mo05b} a phenomenological kinetic model, which
can realize complete FO in a FO layer of finite thickness, $L$, was proposed.
This model allows us to make a link between the two extreme scenarios, discussed above: $L=0$ and $L\rightarrow \infty$.
In fact it is a generalization of a simple kinetic model studied in Refs. \cite{old_SL_FO,old_TL_FO}, which
required $L\rightarrow \infty$ for complete FO.

According to the above references, the basis of the model, i.e. the
invariant escape probability within the FO layer of the thickness $L$, 
for both time-like and space-like normal vectors is given as \cite{kemer,Mo05a,Mo05b}
\be
   P_{esc} =
   \left( \frac{L}{L-x^\mu d\sigma_\mu} \right)
   \left(\frac {p^\mu d\sigma_\mu}{p^\mu u_\mu}\right)\
   \Theta(p^\mu d\sigma_\mu)\,,
\label{esc1}
\ee
where $p^\mu$ is a particle momentum, $u^\mu$ is the flow velocity and $\lambda$ is some characteristic
length scale, e.g. mean free path.
Here we will concentrate on the space-like case only, where the above $\Theta$ function
is important. Calculations will be done in the reference frame of front where space-like normal vector
is $d\sigma^\mu=(0,1,0,0)$.

The two components of the distribution function, $f$: the interacting, $f_i$, and the frozen out,
$f_f$  ones, ($f=f_i+f_f$), 
evolve according to the following equations
(for the details of this 1D model and its applicability
range check Ref. \cite{Mo05a}):
\be
\frac{\partial f_{i}}{\partial x}  = - \frac{P_{esc} }{\lambda}f_{i} + \frac{ f_{eq}(x) - f_{i} }{\lambda_0} \, ,
\quad
\frac{\partial f_{f}}{\partial x}  =  \frac{P_{esc}}{\lambda} f_{i} \, .
\label{sys}
\ee
Here the rethermalization of the interacting component is taken into account via the relaxation time approximation,
where $f_i$ approaches the equilibrated J\"uttner distribution, $f_{eq}(x)$, with a
relaxation length, $\lambda_0$. The system (\ref{sys}) can be solved semi-analytically
in the immediate rethermalization limit \cite{Mo05a}.
The solution is particularly simple for the massless ideal gas without conserved charges.
Below we present results for such a system aiming for a qualitative illustration of the FO process.

\begin{figure}[htb!]
\vspace*{-0.25cm}
                 \insertplot{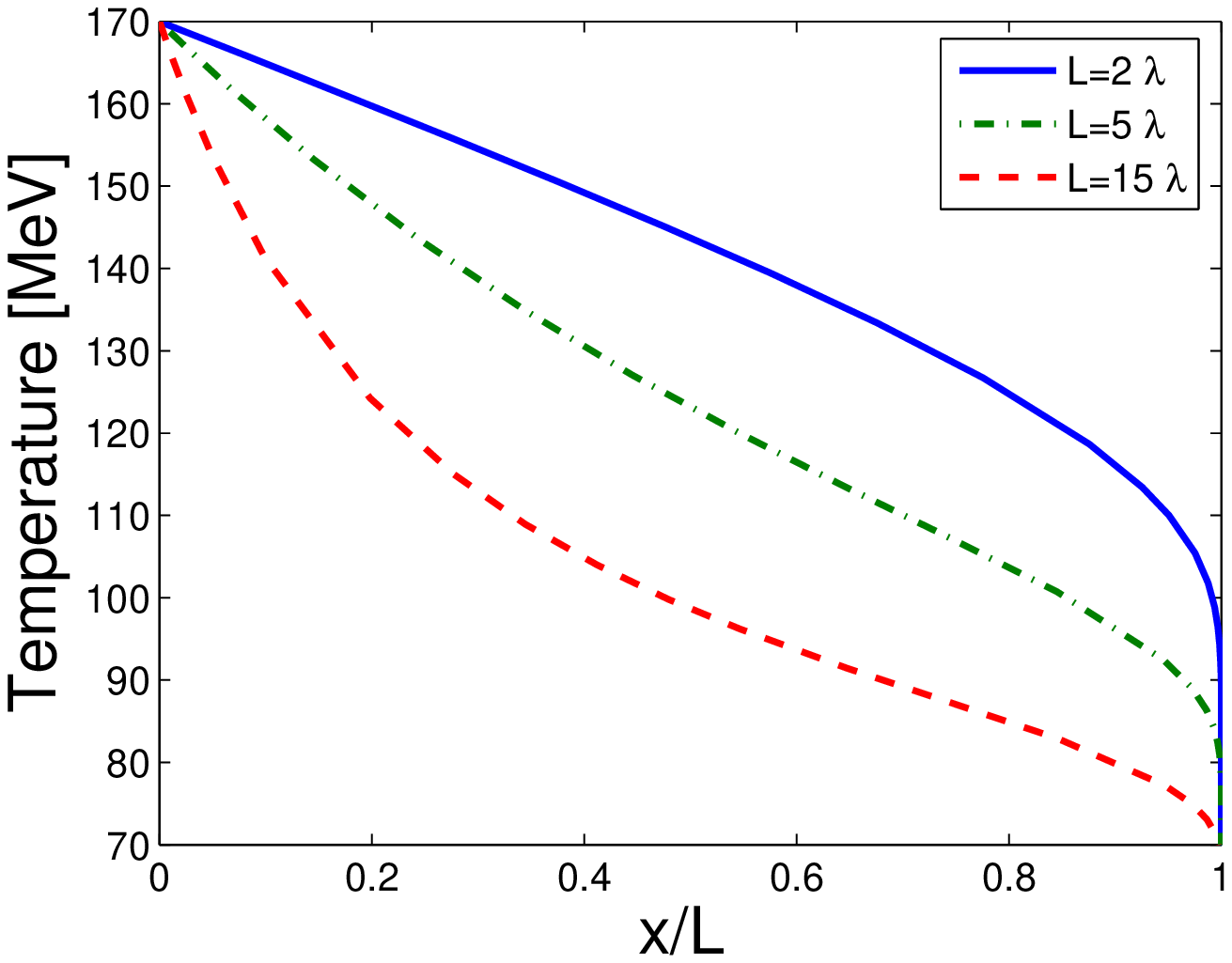}
\vspace*{-0.85cm}
                 \insertplot{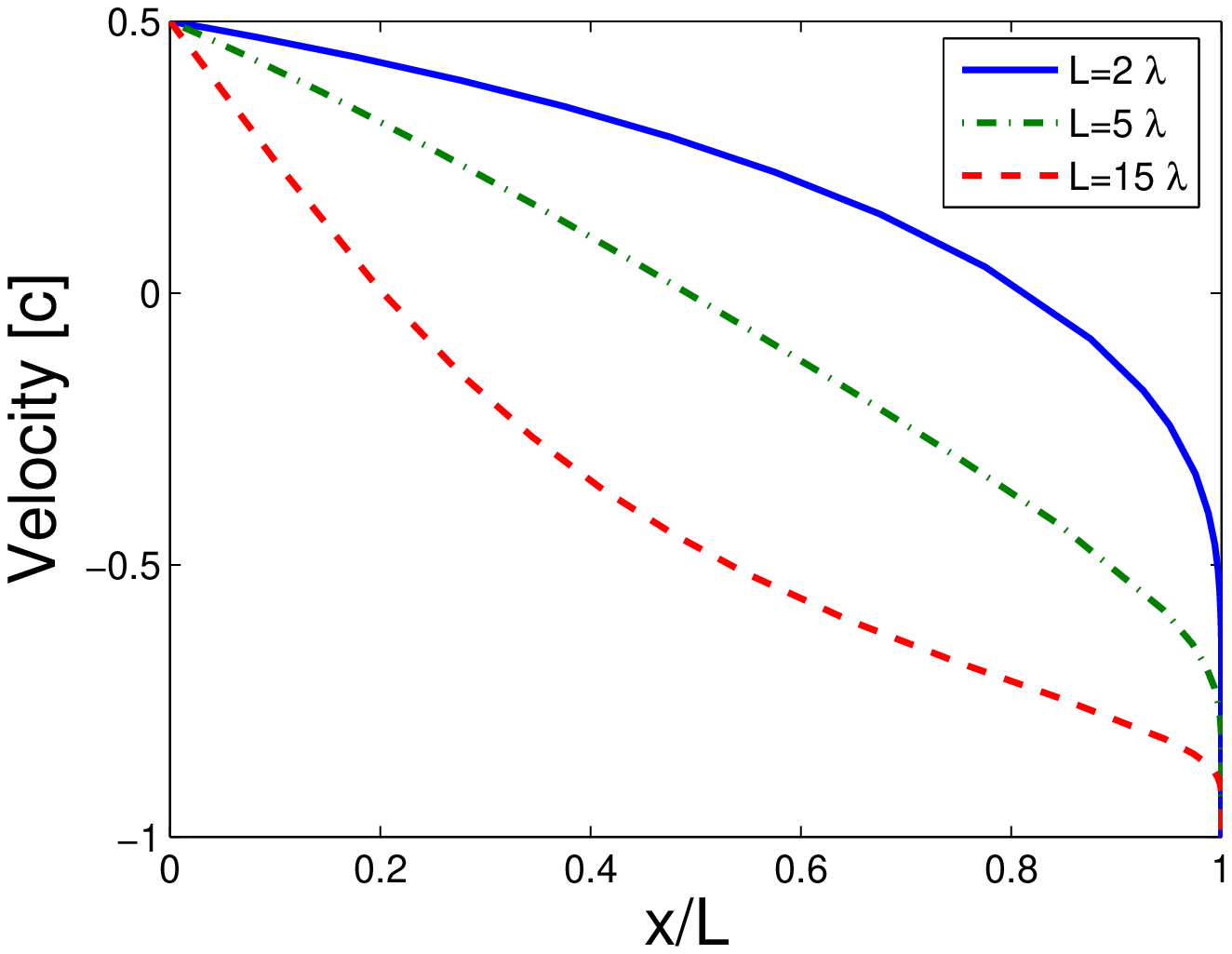}
\vspace*{-1.1cm}
\caption{Evolution of the interacting component: upper subplot shows the evolution of the temperature of the interacting (equilibrated) component for
different FO layers, lower subplot - the evolution of its flow velocity. }
\label{fig1}
\vspace*{-0.5cm}
\end{figure}

\begin{figure}[htb]
\vspace*{-0.25cm}
                 \insertplot{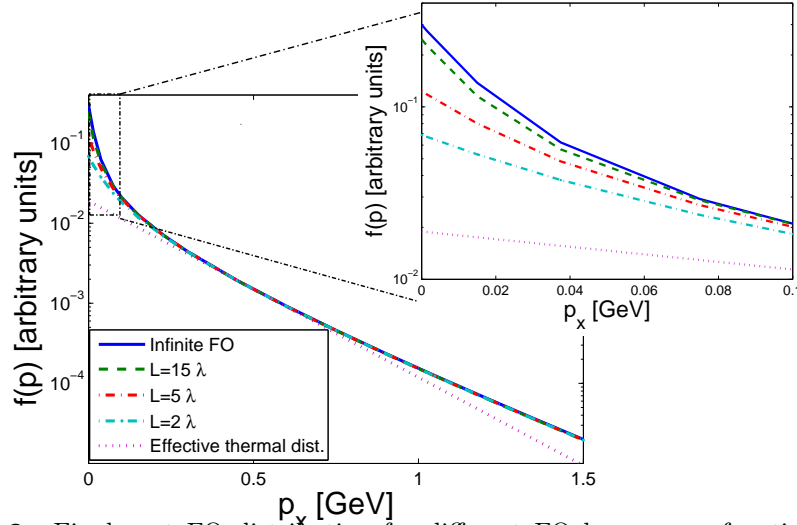}
\vspace*{-1.1cm}
\caption{Final post FO distribution for different FO layers as a function of the momentum in the
FO direction, $p_t=p^x$ in our case ($p^y=p^z=0$). The initial conditions are the same as in
Figs. \ref{fig1}: $T(0)=170\ MeV$, $v(0)=0.5$. The curve for $L\rightarrow \infty $ was calculated in the first
version of the model \cite{old_SL_FO}. The dotted line shows some effective thermal distribution with $T=140\ MeV$ and $v=0.33$. }
\label{fig2}
\vspace*{-0.5cm}
\end{figure}

Figs. \ref{fig1},\ref{fig2} show the results of our calculations.  
As it was already shown in \cite{Mo05a} the final post FO particle distributions
are non-equilibrated distributions, which deviate from thermal ones
particularly in the low momentum region.
By introducing  and varying the thickness of the FO layer, we are strongly affecting the evolution
of the interacting component - see Fig. \ref{fig1}, but for $L>5\lambda$
the final post FO distribution  looks very close to our results for an infinitely long FO calculations,
see \cite{old_SL_FO}.
If we make the thickness of the layer, $L$, very small, e.g. $L=2\lambda$, then
the final post FO distribution is practically a thermal distribution.
Here FO process does not have enough space to distort the thermal shape of the distribution.

We can propose the following algorithm for a simulation where our idealized model would be directly applicable.
1) If  the post FO spectra are curved at low $p_t$ then FO layer is relatively
thick, $L\ge 5 \lambda$, otherwise we have a narrow FO layer, which can be well approximated
by FO hypersurface.
2) If the post FO spectrum (experimental) is curved, then it doesn't matter
how thick was FO layer, we do not need to model the FO dynamics in simulations of the collisions.
Once we have a good parameterization of the post FO spectrum
(still asymmetric, non-thermal), it is enough to write down the conservation laws and non-decreasing entropy
condition with this distribution function \cite{cikk_2} and probably with some volume scaling factor to
effectively account for the expansion during FO.
Thus, our results may justify the use of FO hypersurface in hydrodynamical models for heavy ion collisions,
but with a proper non-thermal post FO distributions.
In large scale simulations for space-like FO the Canceling J\"uttner distribution \cite{karolis} may be a
satisfactory  approximation.



\vfill\eject

\begin{thebibliography}{99}


%
\bibitem{ModifiedBTE}
V.K. Magas et al.,
Nucl. Phys. {\bf A749} (2005) 202;
L.P. Csernai et al.,
hep-ph/0406082; Eur. Phys. J. {\bf  A25} (2005) 65.

\bibitem{hydro_cascade}
D. Teaney et al., Phys. Rev. Lett. {\bf  83} (1999) 4951;
S.A. Bass, A. Dumitru, Phys. Rev. {\bf C61} (2000) 064909;  C. Nonaka, S.A. Bass, nucl-th/0510038,
talk presented at "Quark Matter 2005".

%
\bibitem{Mo05a}
E. Moln\'ar et al.,  nucl-th/0503047. 

\bibitem{Mo05b}
E. Moln\'ar et al.,  nucl-th/0503048.

\bibitem{old_SL_FO}
Cs. Anderlik et al.,
Phys. Rev. {\bf C59} (1999) 388; Phys. Lett. {\bf B459} (1999) 33;
V.K. Magas et al.,
Heavy Ion Phys. {\bf 9} (1999) 193;
Nucl. Phys. {\bf A661} (99) 596.

\bibitem{old_TL_FO}
V.K. Magas et al.,
Eur. Phys. J. {\bf C30} (2003) 255.


\bibitem{kemer}
L.P. Csernai et al.,  hep-ph/0401005; E. Moln\'ar et al.,  nucl-th/0510062, also in these Proceedings.

%
\bibitem{cikk_2}
Cs. Anderlik et al.,
Phys. Rev.  {\bf C59} (1999) 3309.

%
\bibitem{karolis}
K. Tamosiunas and L.P. Csernai,  Eur. Phys. J. {\bf A20} (2004) 269.


\end{thebibliography}
\end{document}